\documentclass[a4paper,fleqn,usenatbib,useAMS]{mnras}

\usepackage{graphicx}	
\usepackage{amsmath}	
\usepackage{amssymb}	
\usepackage{multicol}        
\usepackage{bm}		
\usepackage{pdflscape}	

\newcommand {\be} {\begin{equation}}
\newcommand {\ee} {\end{equation}}

\usepackage[T1]{fontenc}
\usepackage{ae,aecompl}

\usepackage{txfonts}

\title[A spine-sheath model for strong-line blazars]{A spine-sheath model for strong-line blazars}

\author[M. Sikora, M. Rutkowski, M. C. Begelman]{Marek Sikora,$^{1}$\thanks{Contact e-mail: \href{mailto:sikora@camk.edu.pl}{sikora@camk.edu.pl}}
Mieszko Rutkowski,$^{2}$
and Mitchell C. Begelman$^{3,4}$
\\
$^{1}$Nicolaus Copernicus Astronomical Center, Bartycka 18, 00-716 Warsaw, Poland
\\
$^{2}$Marian Smoluchowski Institute of Physics, Jagiellonian University, \L ojasiewicza 11, 30-348 Cracow, Poland
\\
$^{3}$JILA, University of Colorado and National Institute of Standards and Technology, 440 UCB, Boulder, CO 80309, USA
\\
$^{4}$Department of Astrophysical and Planetary Sciences, University of Colorado, 391 UCB, Boulder, CO 80309, USA}

\date{}

\pubyear{2015}

\begin{document}
\label{firstpage}
\pagerange{\pageref{firstpage}--\pageref{lastpage}}
\maketitle

\begin{abstract}
We have developed a quasi-analytical model for the production of radiation in strong-line blazars, assuming a spine-sheath jet structure.
The model allows us to study how the spine and sheath spectral components depend on parameters describing the geometrical and physical structure of ``the blazar zone''.
We show that typical broad-band spectra of strong-line blazars can be reproduced by assuming the magnetization parameter to be of order unity and reconnection to be the dominant
dissipation mechanism. Furthermore, we demonstrate that the spine-sheath model can explain why gamma-ray variations are often observed to have much larger amplitudes than the corresponding
optical variations.  The model is also less demanding of jet power than one-zone models, and can reproduce the basic features of  extreme $\gamma$-ray events.
\end{abstract}




\begingroup
\let\clearpage\relax
\endgroup
\newpage

\section{Introduction}
Given the complexity of AGN jet physics (e.g., \citealt{Begelman1984}), one may be surprised by the success of one-zone, homogeneous models in fitting observed blazar spectra
(see \citealt{Ghisellini2014} and refs. therein). 
In large part, this success is due to the fact that most blazar radiation is emitted in very localized regions.  This is indicated by the time scales of variations and their correlations
between different spectral bands.  In the case of blazars associated with flat-spectrum radio quasars (FSRQs), such correlations are found to be 
particularly strong  between optical and $\gamma$-ray variations (\citep{Bonning2012,Cohen2014,Hovatta2014}). This implies co-spatial production
of low- and high-energy spectral components, the former assumed to be emitted by the synchrotron mechanism, the latter most likely  produced by the external-radiation Compton (ERC) process
(\citealt{2009ApJ...704...38S} and refs. therein).  Sometimes, however, the $\gamma$-ray flux variations --- particularly during higher states --- have much larger fractional amplitudes than the optical
variations \citep{Wehrle1998,Marscher2010,Vercellone2011,Ackermann2014,Cohen2014,Carnerero2015,Hayashida2015}, which is difficult to explain if both spectral components are produced by the
same population of relativistic electrons.  This situation is likely to result from contamination of the optical radiation produced co-spatially with $\gamma$-rays by slowly-varying radiation from another region in the jet.
In order to avoid simultaneous contamination (and suppression) of the $\gamma$-ray variations, the contaminating source in this region should have a lower bulk Lorentz factor $\Gamma$ and/or higher comoving magnetic field $B$ than the bursting region.
(This is because the ERC-to-synchrotron luminosity ratio, $L_{\rm ERC}/L_{\rm syn}$, is proportional to $\Gamma^2/B^2$). We assume that such a source is  associated with slower moving sheaths/layers of laterally stratified jets.

Spine-sheath structures, with a spine moving faster than the sheath, have already been investigated in the context of astrophysical jet phenomenology
\citep{Sol1989,Celotti2001,Ghisellini2005,Tavecchio2008,Darcangelo2009,Mimica2015,Janiak2015}. Their MHD structure was investigated
by \citet{Bogovalov2005}, \citet{Gracia2005}, \citet{Nishikawa2005}, \citet{Mizuno2007}, and \citet{Beskin2006}, and their stability conditions were analyzed by \citet{Hardee2007a} and \citet{Hardee2007b}.
So far, the spine-sheath model has not been worked out in detail for blazars with dense radiative environments, where ERC is the dominant radiation process.
In particular, it has not been determined how significant a contribution Comptonization of synchrotron radiation emitted in the sheath can make to $\gamma$-ray production in the spine.
In this paper we present a model that allows us to make analytic estimates of all spine and sheath spectral components and their dependence on such parameters as powers, magnetizations, opening angles and bulk Lorentz factors.
The spine-sheath model and its basic assumptions are introduced in section~\ref{sec:2}; the approximate formulae allowing calculation of luminosities and frequencies of different spectral components are presented in sections~\ref{sec:2.1} and~\ref{sec:2.2},
respectively; and the dependence of these luminosities and frequencies on the viewing angle are illustrated for an exemplary model in section~\ref{sec:3}. In sections~\ref{sec:4.1}--~\ref{sec:4.4} we discuss how the spine-sheath model can resolve
several problems that afflict one-zone models for spectra and variability of strong-line blazars, and summarize the main results in section~\ref{sec:4.5}.


\section{Spine-sheath model}
\label{sec:2}
\begin{figure}
\begin{center}
\includegraphics[scale=0.5]{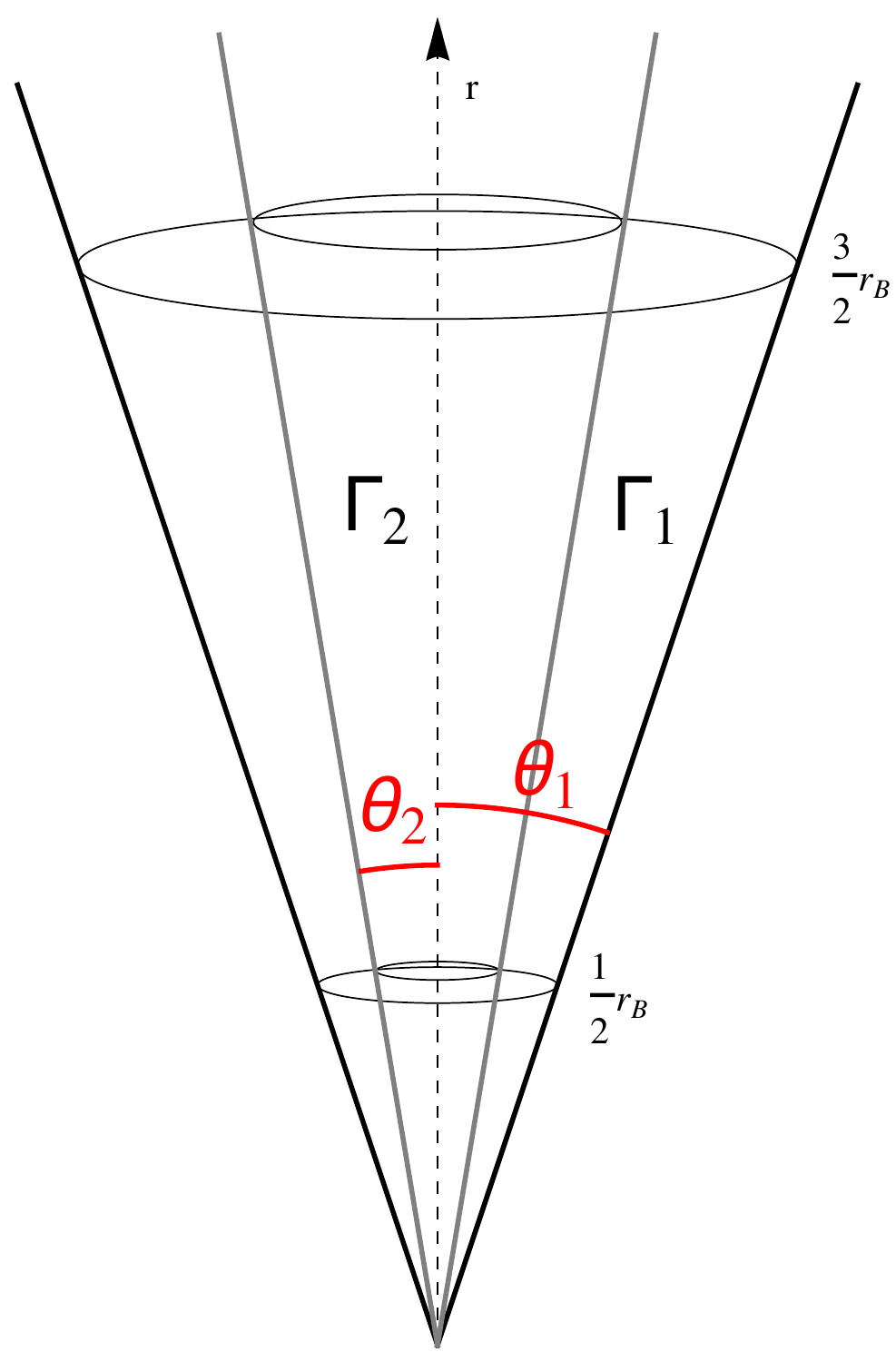}
\caption{Scheme of the spine-sheath structure of a jet.}
\label{fig:structure} 
\end{center}
\end{figure}

We assume that the jet has the ``spine-sheath" conical geometry (see \ref{fig:structure}), in which the Lorentz factor of the spine, $\Gamma_2$, is much larger than that of the sheath, $\Gamma_1$.
The spine and sheath each possess an ``active'' zone, where most of blazar radiation is produced, within the same distance range $r_{\rm bl}/2 < r < 3r_{b\rm l}/2$.
Just prior to the active zone the powers of both jet components are dominated by a sum of the magnetic energy flux, $L_{ B0,i}$, and the kinetic energy flux of cold protons, $L_{\rm p0,\textit{i}}$.
These powers, expressed as a function of the mass flux, $\dot M$, and magnetization, $\sigma_{0,i} \equiv L_{ B0,i}/L_{\rm p0,i}$, are
\be P_{0,i} = L_{ B0,\textit{i}} + L_{\rm p0,\textit{i}} = (1+\sigma_{0,i}) (\Gamma_{0,i}-1) \dot M_{\rm p,\textit{i}} c^2 \, , \label{eq:1}\ee
while magnetic energy fluxes are 
\be L_{ B0,\textit{i}} = \frac {\sigma_{0,i}}{1+\sigma_{0,i}} \, P_{0,i} \, , \ee
where $i=1,2$. We assume that steady-state, axisymmetric jets can dissipate their energy via magnetic reconnection, as well as in reconfinement shocks.  The efficiency of dissipation via these processes is given by
\be \eta_{\rm diss,\textit{i}} = \frac {P_{0,i} - P_i}{P_{0,i}} = 1 - \frac{(1+\sigma_i)(\Gamma_i-1)}{(1+\sigma_{0,i})(\Gamma_{0,i} - 1)}  \label{eq:3} \ee
\citep{Sikora2013}, where $P_i$, $\sigma_i$ and $\Gamma_i$ are values following the dissipation event. 

\subsection{Luminosities}
\label{sec:2.1}
In addition to synchrotron and ERC emission, radiation is produced by the synchrotron-self-Compton (SSC) process and by ``external-synchrotron Compton" (ESC) emission, i.e.,
Comptonization of sychrotron radiation produced in the sheath (spine) by the spine (sheath). Ratios between the luminosities of the various  components can be found from the following approximate relations,
\be L_{\rm{syn,em},\textit{i}}/L_{\rm{ERC,em},\textit{i}} \simeq 
u_{B,\textit{i}}^{(i)}/u_{\rm{ext}}^{(i)} \, , \label{eq:4}\ee
\be L_{\rm{SSC,em},\textit{i}}/L_{\rm{ERC,em},\textit{i}} \simeq 
u_{\rm{syn},\textit{i}}^{(i)}/u_{\rm{ext}}^{(i)} \, , \ee
\be L_{\rm{ESC,em},\textit{i}}/L_{\rm{ERC,em},\textit{i}} \simeq 
u_{\rm{syn},\textit{k}}^{(i)}/u_{\rm{ext}}^{(i)} \, , \ee
where energy densities measured in the ${(i)}$ jet comoving frame are
\be u_{\rm ext}^{(i)} = \frac {\zeta L_{\rm d}}{4 \pi r_{\rm bl}^2 c} \Gamma_i^2 \, , 
\ee
\be u_{B,\textit{i}}^{(i)} = 
\frac {L_{ B,\textit{i}}}{\kappa_{\rm  B\textit{i}} \kappa_{\theta,i} \pi r_{\rm bl}^2 c (\theta_i \Gamma_i)^2} \, , 
\ee
\be u_{\rm syn,\textit{i}}^{(i)} \simeq 
\frac {L_{\rm syn,em,\textit{i}}}{2 \pi r_{\rm bl}^2 c (\theta_i \Gamma_i)} \, , 
\ee
\be u_{\rm syn,\textit{k}}^{(i)} = u_{\rm syn,\textit{k}}^{(k)} \Gamma_{12}^2 \xi_{k}^{(i)} \, , 
\ee
the magnetic energy flux is 
%
\be L_{ B,\textit{i}} = \frac {\sigma_i}{1+\sigma_i} P_i = \frac {\sigma_i} {1 + \sigma_{0,i}} \,  \frac {\Gamma_i - 1}{\Gamma_{0,i}-1} \, P_{0,i} \, , \label{eq:11}\ee
$L_{\rm d}$ is the disk luminosity, $\theta_i$ is the half-opening angle of jet component $i$ (see Fig. \ref{fig:structure}), and the parameters $\zeta$, $\kappa_{ B,\textit{i}}$, $\Gamma_{12}$, and $\xi_k^{(i)}$ are defined as follows:
$\zeta$ is the fraction of the maximal energy density of external radiation as measured in the jet co-moving frame, limited by the geometry and opacity of the matter which reprocesses/isotropizes
the radiation of the accretion disk (see \citealt{Janiak2015} and refs. therein); $\kappa_{B,i}$ is the ratio of the magnetic enthalpy to the magnetic energy density ($\kappa_B=2$ for a jet with magnetic
field dominated by the toroidal component and $\kappa_B=4/3$ for chaotic magnetic fields);
$\kappa_{\theta,2}=1$, $\kappa_{\theta,1}= 1 - (\theta_2/\theta_1)^2$; $\Gamma_{12} = (1 - \beta_1\beta_2) \Gamma_1\Gamma_2 $ ($\sim \Gamma_2/(2 \Gamma_1)$ for $\Gamma_2 \gg \Gamma_1 \gg 1$) is the ``relative" Lorentz factor
between the spine and the sheath; and $\xi_k^{(i)}$ is the factor by which the radiation produced in the $k$-jet is diluted in the     $i$-jet.

Combination of Eqs. \ref{eq:4}-\ref{eq:11} gives
\be l_{\rm syn,em,i} = \frac {4 l_{B,i}}{ \kappa_{ B,\textit{i}} \kappa_{\theta,\textit{i}}\zeta (\theta_i \Gamma_i)^2
\Gamma_i^2 } l_{\rm ERC,em,\textit{i}}\, , \label{eq:12}\ee
\be l_{\rm SSC,em,\textit{i}} = \frac {8 l_{\rm B,\textit{i}}}{ \kappa_{B,\textit{i}} \kappa_{\theta,i} \zeta^2 (\theta_i \Gamma_i)^3\Gamma_i^4 } l_{\rm ERC,em,\textit{i}}^2\, , \ee
\be l_{\rm ESC,em,\textit{i}} = \frac {8 l_{ B,k} \Gamma_{12}^2 \xi_k^{(i)}}{ \kappa_{ B,k} \kappa_{\theta,k} \zeta^2 (\theta_k \Gamma_k)^3 \Gamma_i^2 \Gamma_k^2} \, l_{\rm ERC,em,\textit{i}} l_{\rm ERC,em,\textit{k}}\, , \label{eq:14}\ee
where magnetic energy flux and radiation luminosities are normalized to the disk luminosity ($l_{...} \equiv L_{...}/L_{\rm{d}}$). Fixing the model parameters, one can find all emission luminosities by solving a system
of two quadratic equations with two unknown variables, $l_{\rm ERC,em,\textit{1}}$ and $l_{\rm ERC,em,\textit{2}}$,
\be l_{\rm ERC,em,\textit{i}} + l_{\rm syn,em,\textit{i}} + l_{\rm SSC,em,\textit{i}} + l_{\rm ESC,em,\textit{i}} = \epsilon_i p_{0,i}\, , \ee
where $p_{0,i} = P_{0,i}/L_d$, and $\epsilon_i$ is the fraction of the $i$-jet power converted to radiation.

Having computed the emitted luminosities, one can calculate the observed luminosities as a function of the angle between the jet axis and the direction to the observer, $\theta_{\rm obs}$,
\be l_{\rm ERC,\textit{i}} = {\cal A}_{ERC,\textit{i}} \, l_{\rm ERC,em,\textit{i}} \, , \ee 
and all remaining spectral components are amplified by a factor ${\cal A}_{iso,i}$, where the luminosity amplification functions ${\cal A}_{ERC,i}$ and ${\cal A}_{iso,i}$ are given by Eqs. \ref{eq:A1} and \ref{eq:A2}. For jet opening angles 
$\theta_i \ll 1/\Gamma_i$  they can be approximated by 
\be {\cal A}_{ERC,i} = ({\cal D}_i/\Gamma_i)^5 \, \Gamma_i^2  \, , \ee
\be {\cal A}_{iso,i}  = ({\cal D}_i/\Gamma_i)^3 \, \Gamma_i^2 \, , \ee
where ${\cal D}_i = [\Gamma_i(1-\beta_i \cos{\theta_{\rm obs}}]^{-1}$. Accuracy of these approximations is illustrated in Fig. \ref{fig:amp-approximation}.  The different amplification function for ERC luminosity comes from the fact that the
ERC radiation in the jet comoving frame is not isotropic \citep{Dermer1995}.  ESC radiation is also anisotropic in the comoving frame, however, because the relative velocity of the spine and sheath flows is only mildly relativistic,
this anisotropy is weak and its effect on the luminosity amplification can be ignored.

\subsection{Frequencies}
\label{sec:2.2}
Assuming that each electron is accelerated once, the injection rate of accelerated electrons is equal to the electron number flux.  This gives 
\be \dot N_{\rm e,\textit{i}} \bar E_{\rm e,\textit{i}} = 
\eta_{\rm e,\textit{i}} \eta_{\rm diss,\textit{i}} P_{0,i} \, , \ee
where $\dot N_{\rm e,\textit{i}}$ is the  electron number flux, $\bar E_{\rm e,\textit{i}}$ is the average energy of the accelerated electrons, and $\eta_{\rm e,\textit{i}}$ is the fraction of dissipated energy tapped by electrons. Since 
\be \bar E_{\rm e,\textit{i}} = \bar E_{\rm e,\textit{i}}^{(i)} \Gamma_i = (\bar \gamma_i - 1) m_{\rm e}c^2 \Gamma_i \ , \ee
we obtain the average Lorentz factor of accelerated electrons
\be \bar \gamma_i = \frac {\eta_{\rm e,\textit{i}} \eta_{\rm diss,\textit{i}} P_{0,i}}{m_{\rm e}c^2 \dot N_{\rm e,\textit{i}} \Gamma_i} +1 \, . \ee
Inserting here Eq. \ref{eq:1} with $\dot M_{\rm p,\textit{i}}= \dot N_{\rm p,\textit{i}} m_{\rm{p}} c^2$ and $\Gamma_{0,i} \gg 1$ gives
\be \bar \gamma_i = \eta_{\rm e,\textit{i}} \eta_{\rm diss,\textit{i}} {m_{\rm p}n_{\rm p}\over m_{\rm e}n_{\rm e}}(1+\sigma_{0,i}){\Gamma_{0,i}\over\Gamma_i} +1 \, . \label{eq:22}\ee
Assuming a broken-power-law electron injection function with a break at $\bar \gamma_i$, one can find that the peaks of different spectral components in $\nu L_{\nu}$ are at frequencies \citep{2015MNRAS.449..431J}
\be \nu_{\rm ERC,\textit{i}} \simeq \bar \gamma_i^2 {\cal D}_i^2 \nu_{\rm ext} \, , \ee
where $\nu_{ext}$ is the average frequency of the dominant external radiation 
field,
\be \nu_{\rm syn,\textit{i}} \simeq 3.7 \times 10^6 \bar \gamma_i^2 B_i^{(i)} {\cal D}_i
\, , \ee
\be \nu_{\rm SSC,\textit{i}} \simeq \bar \gamma_i^2 \nu_{\rm syn,\textit{i}} \, , \ee
\be \nu_{\rm ESC,\textit{i}} \simeq \bar \gamma_i^2 \nu_{\rm syn,\textit{k}}^{(i)} {\cal D}_i = \bar \gamma_i^2 \Gamma_{12} \nu_{\rm syn,\textit{k}}^{(k)} {\cal D}_i = \bar \gamma_i^2 \Gamma_{12} \nu_{\rm syn,\textit{k}} ({\cal D}_i/{\cal D}_k) \, , \ee
where $B_i^{(i)} = \sqrt{8\pi u_{B,i}^{(i)}}$, and
\be u_{B,i}^{(i)} = 
\frac {l_{B,i}} {\kappa_{B,i}\kappa_{\theta_i} \pi (r_{\rm bl}^2/L_{\rm d})
 c (\theta_i \Gamma_i)^2 } \, . \label{eq:27}\ee

\section{Results}
\label{sec:3}
\subsection{Exemplary model}

For the purpose of presenting a representative model we will assume that: 
1)~prior to the dissipation zone the power of the jet per solid angle is constant across the jet; 2) the half-opening angle of the spine is equal to the Doppler angle (i.e. $\theta_2\Gamma_2=1$); and 3) the energy
dissipation is dominated by magnetic reconnection. The latter assumption allows us to use the approximation $\Gamma_i=\Gamma_{0,i}$, whence Eq. \ref{eq:11} gives
\be l_{B,i} =  \frac {\sigma_i}{1+ \sigma_{0,i}} p_{0,i} \, , \label{eq:28}\ee
and from Eq. \ref{eq:3}
\be \eta_{diss,i} = \frac {\sigma_{0,i} - \sigma_i} {1+ \sigma_{0,i}} \, . \ee

\begin{figure}
  \begin{center}
    \includegraphics[width=.9\columnwidth, angle=0]{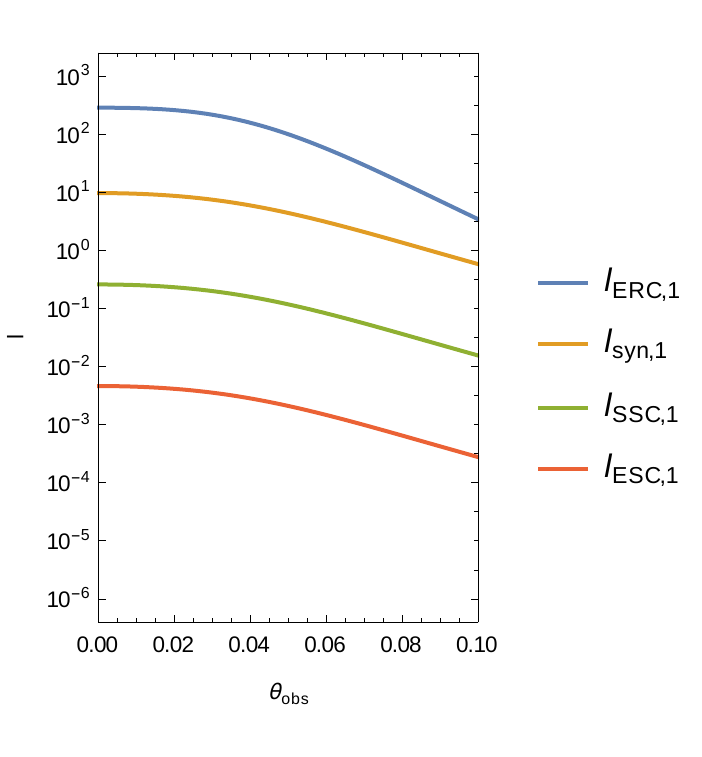}
    \caption{Luminosities normalized by $L_d$ vs. the observer angle: sheath.}
    \label{fig:lum-sep1} 
  \end{center}
\end{figure}

\begin{figure}
  \begin{center}
    \includegraphics[width=.9\columnwidth, angle=0]{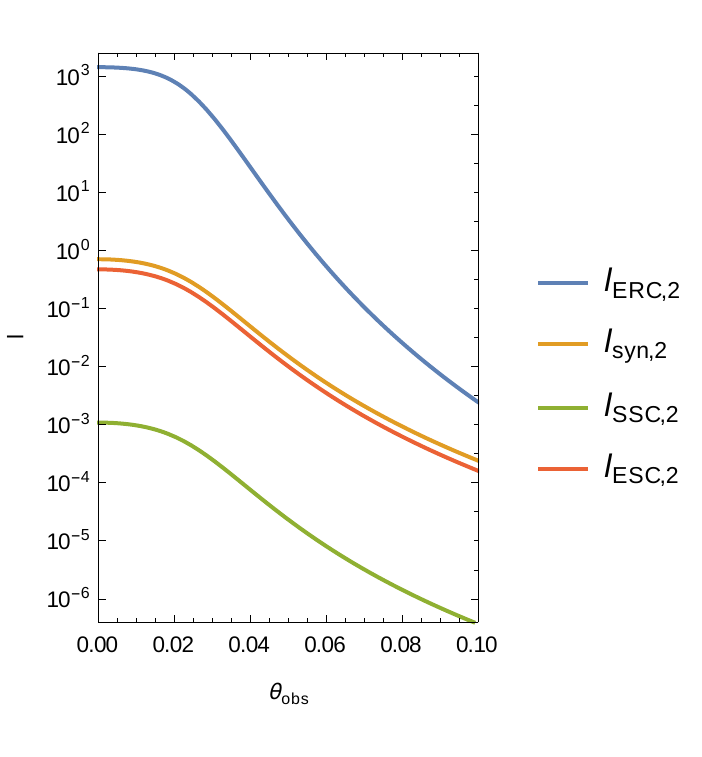}
    \caption{Luminosities normalized by $L_d$ vs. the observer angle: spine.}
    \label{fig:lum-sep2} 
  \end{center}
\end{figure}

\begin{figure}
  \begin{center}
    \includegraphics[width=.9\columnwidth]{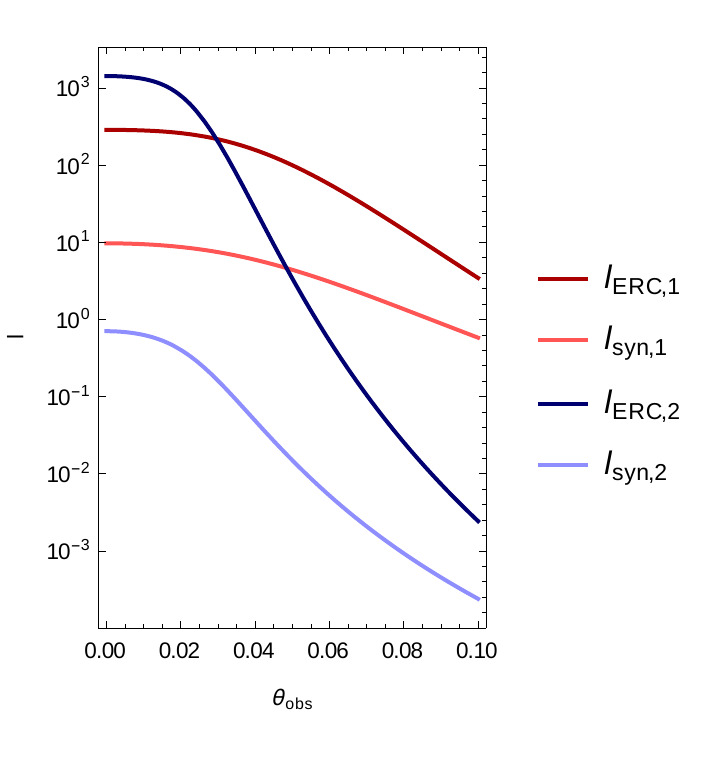}
    \caption{ERC and synchrotron luminosities vs. the observer angle.}
    \label{fig:lum-tog} 
  \end{center}
\end{figure}

Figs. \ref{fig:lum-sep1}, \ref{fig:lum-sep2} and \ref{fig:lum-tog} show the apparent luminosities as a function of the inclination
 angle, calculated for the following set of parameters: 
$p_{0,1}=p_{0,2}=0.5$ ($\to (P_0 = P_{0,1}+P_{0,2})/L_d=1$); $\Gamma_1=12$; $\Gamma_2=48$; $\sigma_{0,1}=\sigma_{0,2}=1$; $\sigma_{1}=\sigma_{2}=0.25$; $\kappa_{B,1}=\kappa_{B,2}=5/3$; $\zeta= 0.03$; $\eta_{\rm e}=0.5$; and $\eta_{\rm rad}=1$ (the latter being
confirmed {\it a posteriori} to be close to unity).  In this case $\epsilon_i \simeq \eta_{rad} \eta_{\rm e} \eta _{\rm diss,\textit{i}} \simeq 3/16$.  
Note that for $p_{0,1}=p_{0,2}$ and $\theta_2\Gamma_2=1$, the assumed 
proportionalities $p_{0,1} \propto \Omega_1 \propto (\theta_1^2-\theta_2^2)$ and $p_{0.2} \propto \Omega_2 \propto \theta_2^2$ imply $\theta_1= \sqrt{2}/\Gamma_2$, i.e. that $\theta_1\Gamma_1 = \sqrt 2 (\Gamma_1/\Gamma_2)$.  
Results presented in Figs. \ref{fig:lum-sep1}, \ref{fig:lum-sep2} and \ref{fig:lum-tog} are obtained using exact formulas for 
the luminosity amplification factors (see Appendix A).
In Fig. \ref{fig:lum-sep1} the dependence of luminosities of the spectral 
components produced in the sheath  on the observed angle are shown. 
Since for the sheath the Doppler angle is $1/\Gamma_1 \sim 0.08$, that 
dependence up to $\theta_{obs} \sim 5^{\circ}$ is pretty modest. We can see 
a clear 
luminosity hierarchy, $l_{ESC,1} \ll l_{SSC,1} \ll l_{syn,1} \ll l_{ERC,1}$, which 
indicates that Comptonization of the synchrotron luminosity produced in 
the spine, $l_{ESC,1}$, cannot compete with other spectral components 
at any frequencies. 
In Fig. \ref{fig:lum-sep2} the dependence of the spine luminosities 
on the observed angle are shown. Here the luminosities of all spectral 
components
drop much faster with increasing observed angle; and ERC, SYN, and SSC
luminosities are much more separated than the respective sheath luminosities. 
Another difference is that $l_{ESC,2}$ is much larger than $l_{SSC,2}$,
and at $\theta_{obs} \sim 1/\Gamma_2$ it can become of order $l_{SSC,1}$.
In  Fig. \ref{fig:lum-tog} we compare ERC and SYN luminosities produced in the spine and 
sheath. Clearly the spine synchrotron luminosity is never important:
at $\theta_{obs} < 1/\Gamma_2$ the ERC radiation is strongly dominated by 
the spine and the synchrotron one by the sheath, while at 
$\theta_{obs} > 3/\Gamma_2$ both ERC and synchrotron components are dominated
by the sheath.

The dependence of apparent luminosities on  model parameters can be easily 
followed by using the simple analytical approximations obtained under the condition
that $l_{ERC,em,i} \gg l_{syn,em,i} + L_{SSC,em,i} + l_{ESC,em,i}$.  This condition 
implies $l_{ERC,em,i} \simeq \epsilon_i p_{0,i}$, and then inserting Eq. \ref{eq:11} into
Eqs. \ref{eq:12}--\ref{eq:14} and multiplying the emitted luminosities by the amplification
functions we obtain:
\be l_{\rm ERC,\textit{i}} \simeq  
\epsilon_i f_i \,\, p_0 \, {\cal A}_{\rm ERC, \textit{i}} \, , \ee
\be l_{\rm syn,\textit{i}} \simeq 4 \,  \frac {c_{\sigma,i} \epsilon_i f_i^2}{\kappa_{B,i}\kappa_{\theta,i} \zeta (\theta_i\Gamma_i)^2} \,\, 
\frac {p_0^2}{\Gamma_i^2} \, {\cal A}_{\rm iso,\textit{i}} \, , \ee
\be l_{\rm SSC,\textit{i}} \simeq 8 \, \frac {c_{\sigma,i} \epsilon_i^2 f_i^3}{\kappa_{B,i}\kappa_{\theta,i} \zeta^2 (\theta_i\Gamma_i)^3} \,\, \frac{p_0^3}{\Gamma_i^4} \,
{\cal A}_{\rm iso,\textit{i}} \, , \ee
\be l_{\rm ESC,\textit{i}} \simeq 8 \, \frac {c_{\sigma,k} \epsilon_i \epsilon_k f_i f_k^2 \Gamma_{12}^2 \xi_k^{(i)}} {\kappa_{B,k} \kappa_{\theta,k} \zeta^2 (\theta_k\Gamma_k)^3} \,\, \frac{p_0^3}{\Gamma_i^2 \Gamma_k^2} \, {\cal A}_{\rm iso,\textit{i}}   \, , \ee
where  $f_i = P_{0,i}/P_0$, $c_{\sigma,i} = \sigma_i/(1+\sigma_{0,i})$, and $\theta_1/\theta_2 =f_2^{-1/2}$ as imposed by the condition $P_1/\Omega_1= P_2/\Omega_2$.

\begin{figure}
  \begin{center}
    \includegraphics[width=\columnwidth, angle=0]{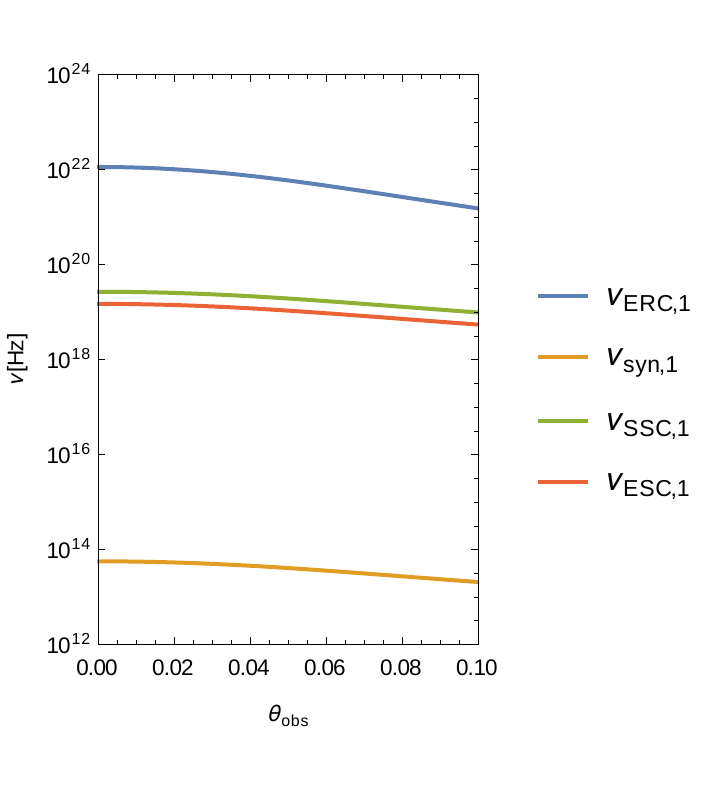}
    \caption{Frequencies of the spectral peaks vs. the observer angle: sheath.}
    \label{fig:frequencies1} 
  \end{center}
\end{figure}

\begin{figure}
  \begin{center}
    \includegraphics[width=\columnwidth, angle=0]{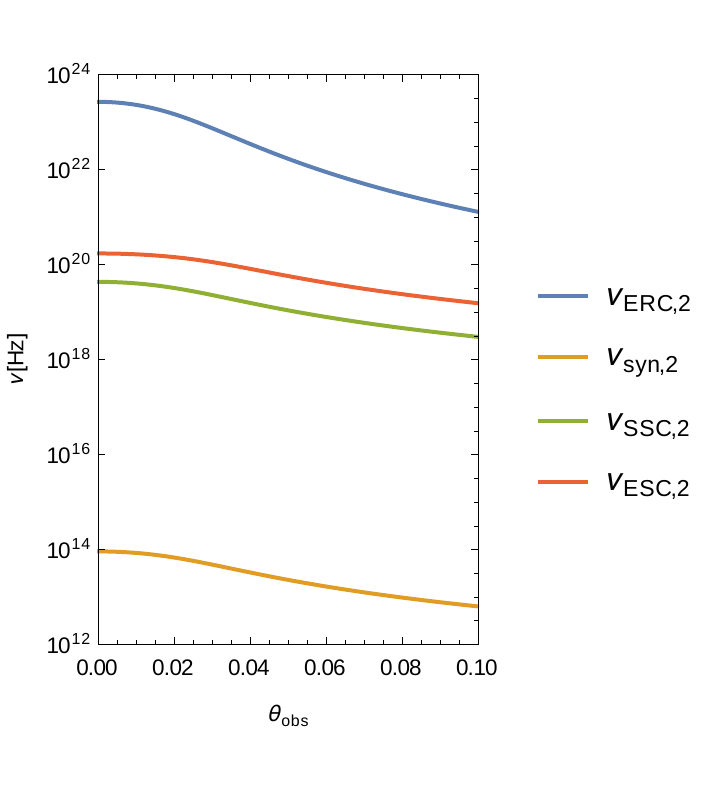}
    \caption{Frequencies of the spectral peaks vs. the observer angle: spine.}
    \label{fig:frequencies2} 
  \end{center}
\end{figure}

The dependence of spectral-peak frequencies of different radiation components on the inclination angle is shown in Figs. \ref{fig:frequencies1} and \ref{fig:frequencies2}.
The peak frequencies  are calculated assuming $n_{\rm e}=n_{\rm p}$, $\nu_{\rm ext} = 1.45 \times 10^{14}$Hz (${\rm h}\nu_{ext}=0.6$eV), $r_{\rm bl} = 0.3 r_{\rm sub}$,
and the sublimation radius for graphite, $r_{\rm sub} \simeq 1.6 \times 10^{-5} \sqrt{L_d}$ (\citealt{Sikora2013} and refs. therein).  For the parameters adopted in our exemplary model, Eq. \ref{eq:22} gives 
$\bar \gamma_i = 690$, and Eqs. \ref{eq:27} and \ref{eq:28} give $u_{B,1}^{(1)} = 0.16 \, {\rm erg\, cm^{-3}}$ ($B_1^{(1)}= 2.0 \, {\rm G}$) and $u_{B,2}^{(2)} = 0.017 \, {\rm erg\, cm^{-3}}$ ($B_2^{(2)}= 0.66 \, {\rm G}$). 

\begin{figure}
  \begin{center}
    \includegraphics[width=\columnwidth, angle=0]{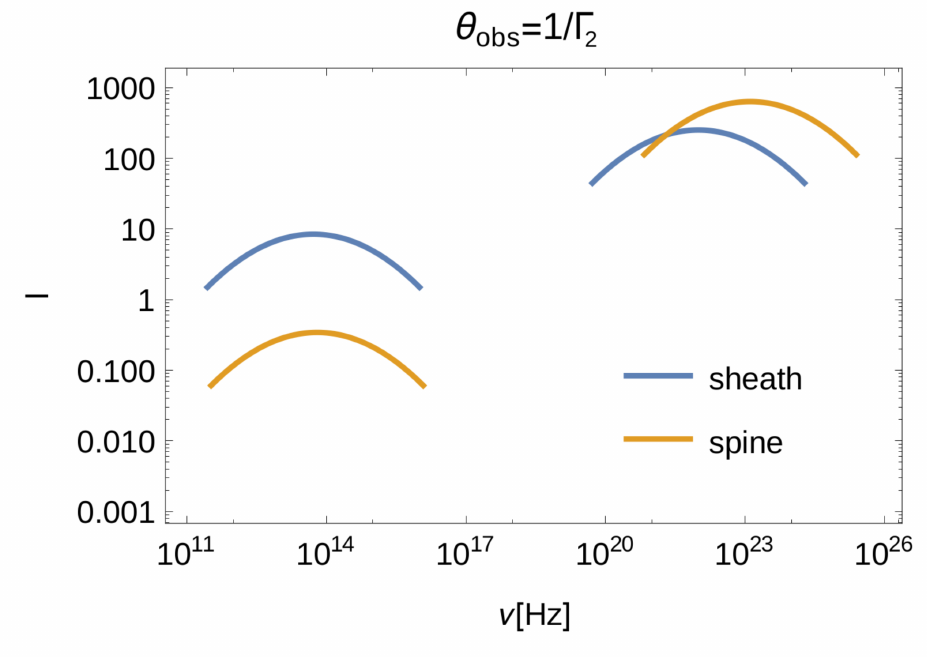}
    \caption{The spine and sheath ERC+synchrotron spectra observed at $\theta_{\rm obs}=1/\Gamma_2$.}
    \label{fig:broad-band} 
  \end{center}
\end{figure}

The schematic broad-band SYN-ERC spectra obtained using results of our exemplary
models for $\theta_{obs} = 1/\Gamma_2$ are illustrated in Fig. \ref{fig:broad-band}.
This figure illustrates that at very low inclination angles the high energy
spectral component is dominated by the ERC radiation produced by the spine, 
while the low energy spectral components is dominated by synchrotron radiation
produced by the sheath.

\section{Discussion and Conclusions}
\label{sec:4}
\subsection{Contamination of optical variations}
\label{sec:4.1}
Lower fractional amplitudes of optical flux variations 
compared to $\gamma$-ray variations, often observed, may result from 
contamination of variable spine radiation by steady-state or weakly variable 
sheath radiation.  As  can be deduced from Fig. \ref{fig:lum-tog}, such an effect is 
expected to be particularly strong for $\theta_{\rm obs} < 1/\Gamma_2$, for which
the observed $\gamma$-rays are strongly dominated by the spine.  


\subsection{$\gamma$-ray outbursts, their time scales and spectra} 
\label{sec:4.2}
The production of powerful $\gamma$-ray outbursts can result from a temporal 
decrease of $\theta_{\rm obs}$ due to a change of the spine direction (here,
$\theta_{obs}$ is defined as measured relative to the instantaneous axis of 
the wiggling spine). Changes of the spine direction may be caused by
current driven instabilities (\citealt{Nalewajko2012}, 
\citealt{Janiak2015}, and refs. therein),
or variations of the jet injection direction due to non-axisymetric interchange 
instabilities which drive  accretion onto the BH in 
``magnetically--arrested--discs'' \citep{McKinney2012}. In the former case 
the time scale of the observed outburst is 
\be t_{\gamma,flare} \sim r_{bl} \Delta \theta_{\rm obs}  / ({\cal D}_2 v_A) \, \ee
where $\Delta \theta_{\rm obs}$ is the change of the spine direction and 
$v_A/c \simeq \sqrt{\sigma_{0,2}/(1+\sigma_{0,2})}$ is the Alfv\'enic velocity.
For our exemplary model parameters this gives 
$\Delta t_{\gamma} \sim 5 (\Delta \theta_{\rm obs}/\theta_2)$ days.
Hence, a  decrease of $\theta_{\rm obs}$ from $1.5 \theta_2$
to $\theta_2$ is followed by an increase in the observed $\gamma$-ray flux  by a factor 6 
(see Fig. \ref{fig:amplification}) with a growth time scale of $\sim 2.5$ days.
In the latter case (variation of the jet injection direction),  no local 
causality constraints apply. Obviously, 
in the proposed spine-sheath model the observed variability patterns are
predicted to be superposed from variations of intrinsic $\gamma$-ray production 
and variations resulting from jet wiggling.
 
In the proposed spine-sheath model, often-seen hardening of the $\gamma$-ray spectrum 
during  powerful outbursts results from the fact that 
the ERC peak of the spine component is produced at higher frequencies than 
the ERC peak of the sheath component (see Fig. \ref{fig:broad-band}).

\subsection {Compton dominance}
\label{sec:4.3}
For observers located at $\theta_{obs} < 1/\Gamma_2$,  the high energy component is predicted to dominate over the low-energy component by a factor 
\be q
\simeq \frac{l_{\rm ERC,2}}{l_{\rm syn,1}}  =\frac{1}{4} \,
\frac {\epsilon_2 f_2 \kappa_{B,1} [1-(\theta_2/\theta_1)^2] \zeta (\theta_1 \Gamma_1)^2} {c_{\sigma,1} \epsilon_1 f_1^2} \, \frac {\Gamma_1^2}{p_0} \,  \frac{{\cal A}_{\rm ERC,2}}{{\cal A}_{\rm iso,1}} \, \ee
which for the parameters of our exemplary model presented in \S3 and an observer located at $\theta_{obs} < 1^{\circ}$ is $\sim 100$ (see Fig. \ref{fig:lum-tog}). For observers located at $\theta_{obs} > 2/\Gamma_2$, our exemplary model predicts
\be q \simeq \frac{l_{\rm ERC,1}}{l_{\rm syn,1}}  = \frac{1}{4} \,\frac {\kappa_{B,1} [1-(\theta_2/\theta_1)^2] \zeta (\theta_1 \Gamma_1)^2} {c_{\sigma,1} f_1} \, \frac {\Gamma_1^2}{p_0} \, \frac{{\cal A}_{\rm ERC,1}}{{\cal A}_{\rm iso,1}} \, , \ee
and for $\theta_{\rm obs} > 2^{\circ}$ gives $q < 10$.  Since most FSRQs have $q<10$ \citep{Giommi2012,Finke2013,Lister2015}, most of them are likely to be observed at $\theta_{\rm obs} > 2^{\circ}$.
However, the level of Compton dominance is found to vary in time within individual objects, reaching the largest values during luminous outbursts (see, e.g., \citealt{Bonnoli2011}; \citealt{Ackermann2014}; \citealt{Hayashida2015}). 



\subsection{Jet energetics}
\label{sec:4.4}
Spine-sheath models are also advantageous for jet energetics.  According to one-zone models, the observed $\gamma$-ray outbursts with luminosities $\sim 10^3 L_d$ require jet powers to be $\sim 100$ times larger than the disk luminosity
\citep{Ghisellini2014}.  In a spine-sheath model with the observer located at $\theta_{obs} < 2^{\circ}$, similar $\gamma$-ray luminosities are possible with $P_0 \sim L_d$ (see Fig. \ref{fig:lum-tog}).
Such lower jet energies are  supported  by observations of blazar radio-halos (\citealt{Kharb2010}; \citealt{Meyer2011}).


\subsection{Conclusions}
\label{sec:4.5}
Transverse structure in AGN jets, with faster spines/cores and slower sheaths/layers, is indicated
by a number of independent observations. On large, $\sim 100$ kpc scales of powerful double radio
sources it is indicated by the fact that
jet velocities, inferred from their sideness flux ratios,
are barely relativistic ($\Gamma \sim 1.2 - 1.4$: \citealt{1997MNRAS.286..425W}; 
\citealt{2009MNRAS.398.1989M}),
while X-ray observations suggest 
$\Gamma \sim 4-14$ \citep{2015IAUS..313..219S}. 
The transverse kinematic structure 
of powerful large scale jets is also required to  explain details
of radio-optical-X-ray spectra of the jet knots \citep{2006ApJ...648..900J}.
On parsec and smaller scales, transverse structure is strongly
indicated by recent mm-VLBI observations of  a jet and a counter-jet  in Cyg A
\citep{Boccardi2015}. These observations show that the lateral width of the jet is much broader 
than typical values deduced from blazar studies and that the flow speed drops in the outer layers 
to mildly relativistic values. Low-power 
jets, 
associated with BL Lac objects and their hosts, Fanaroff-Riley Class I (FRI) radio galaxies, also presumably have 
transverse structure. This is deduced: from edge-brightening  of some parsec-scale jets
(\citealt{2004ApJ...600..127G}; \citealt{2014ApJ...785...53N});
from FRI - BL Lac unification studies \citep{2002A&A...383..104C};
and from mildly relativistic 
speeds deduced from VLBI observations of jets in TeV-BL Lac objects, 
as contrasted with $\Gamma \gg 1$ deduced from modeling their broad-band spectra 
(\citealt{2000A&A...358..104C}; \citealt{2014ApJ...797...25P}).
Additionally, a
spine-sheath structure in these objects obviates the need for  extremely weak magnetic fields that  
are implied if one  assumes transversely uniform jets \citep{2015arXiv150908710T}

In this paper the  spine-sheath jet model is applied to strong-line blazars.
We demonstrate that it provides a natural explanation for a number of observed
features of $\gamma$-ray flares in these objects, including
their hardness, their extreme Compton dominance (compared to the optical flux during a flare),
and the fact that fractional flux variations in the optical band are often observed to be much 
weaker than those in $\gamma$-rays. 
And  contrary to transversely uniform jets, the production of flares  
with a large Compton dominance does not require very low  jet 
magnetization.
However, the model should be treated as a very crude approximation of a real
jet in which the lateral distribution of the Lorentz factor is likely to be a smooth function of angle rather than two-value step-function. This is indicated by recent direct observations of a sub-parsec jet and counter-jet in Cyg A in the mm band \citep{Boccardi2015}.
These observations show that the lateral width of the jet is much broader than typical values deduced from blazar studies and that the flow speed drops in the outer layers to mildly relativistic values.

While the Lorentz factors chosen for our exemplary model (\S3) were motivated by observations,  the relative locations of the $\gamma$-ray and synchrotron components in
the $\log {\nu L_{\nu}}$ --- $\log {\nu}$ plane (Fig. \ref{fig:broad-band}) --- also consistent with observations ---  are determined through physical considerations, by connecting average electron energies with the efficiency of the reconnection process,
assuming energetic coupling between electrons and protons, and taking the magnetization parameter $\sigma$ to be of order unity, as motivated by several theoretical studies (see, e.g., \citealt{Komissarov2009}; \citealt{Lyubarsky2010}).

%



An advantage of our toy model is that it allows one to follow analytically the dependence of different radiation spectral components on the various parameters. 
In particular, our exemplary  model assumptions 
that $P_0/\Omega=const$ and that values of $\sigma$ and $\eta_{diss}$ in the 
spine and sheath are the same, are rather conservative. Relaxing  these 
assumptions by allowing lower $\sigma$ and  larger $\eta_{diss}$ in the spine 
and/or
larger power of the jet in the spine per solid angle may help to explain such 
extreme 
events as the one recorded in December 2013 in 3C279 \citep{Hayashida2015}. 

Alternative scenarios to explain extreme $\gamma$-ray outbursts
and their properties have been  proposed recently by 
\citet{Asano2015}
and \citet{Tavani2015}. 
\citeauthor{Asano2015}
consider second--order Fermi 
acceleration within negligibly magnetized plasma ($u_B'/u_e' < 10^{-4}$). Such 
a low magnetization seems to contradict theoretical works showing
the difficulty of converting  initially Poynting flux--dominated flows to
$\sigma \ll 1$ jets \citep{Komissarov2009,Lyubarsky2010}.
\citeauthor{Tavani2015} invoke the mirror model \citep{Ghisellini1996}, assuming  
that cold mirrors/plasmoids can be formed in the jet, following tearing 
instabilities in the reconnection layers. However, getting them cold
requires very efficient cooling, which is rather difficult to obtain
in a jet at thousands of gravitational radii from the BH.

\section*{Acknowledgements}
\addcontentsline{toc}{section}{Acknowledgements}
The research was supported in part by the Polish NCN grant 
UMO-2013/08/A/ST9/00795.
MCB acknowledges support from NSF grant AST-1411879 and NASA Astrophysics
Theory Program grant NNX14AB37G. 







\appendix
\section{Luminosity amplification}
\be {\cal A}_{\rm ERC,\textit{i}} = \Gamma_i^2 \, \frac{1}{\Omega_i} \int\displaylimits_{\phi} \int_{\vartheta_2}^{\vartheta_1} {(\delta_i/\Gamma_i)^5 \, d\Omega} \, , \label{eq:A1}\ee 
\be {\cal A}_{\rm iso,\textit{i}} = \Gamma_i^2 \, \frac{1}{\Omega_i} \int\displaylimits_{\phi} \int_{0}^{\vartheta_2} {(\delta_i/\Gamma_i)^3 \, d\Omega} \, , \label{eq:A2}\ee
where $\delta_i =\frac{1}{\Gamma_i (1-\beta_i\cos{\vartheta})}$, $d\Omega=d(\cos{\vartheta}) \, d\phi$ and ($\vartheta$, $\phi$) are spherical coordinates with z-axis parallel to the direction to the oberver. By $\Omega_i$ we denote total solid angles of jets.

\begin{figure}
  \begin{center}
    \includegraphics[width=\columnwidth]{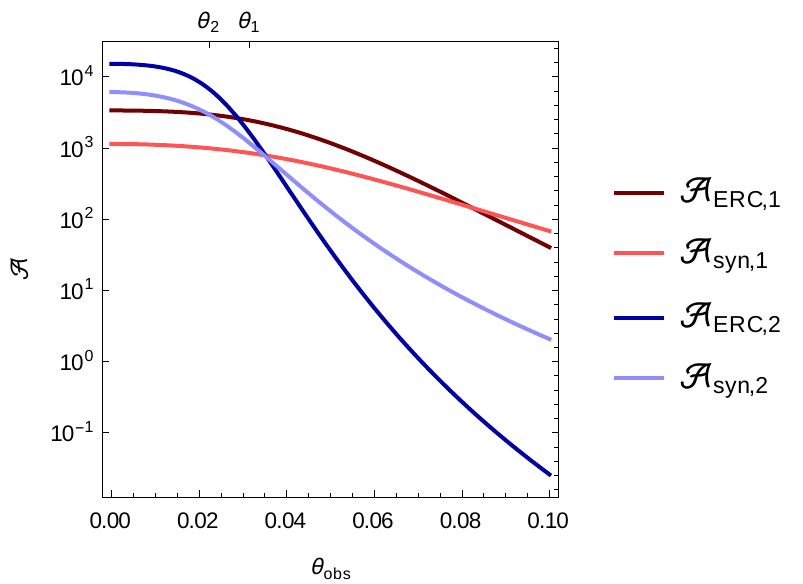}
    \caption{Amplification factors as a function of an observation angle.}
    \label{fig:amplification} 
  \end{center}
\end{figure}

\begin{figure}
  \begin{center}
    \includegraphics[width=\columnwidth]{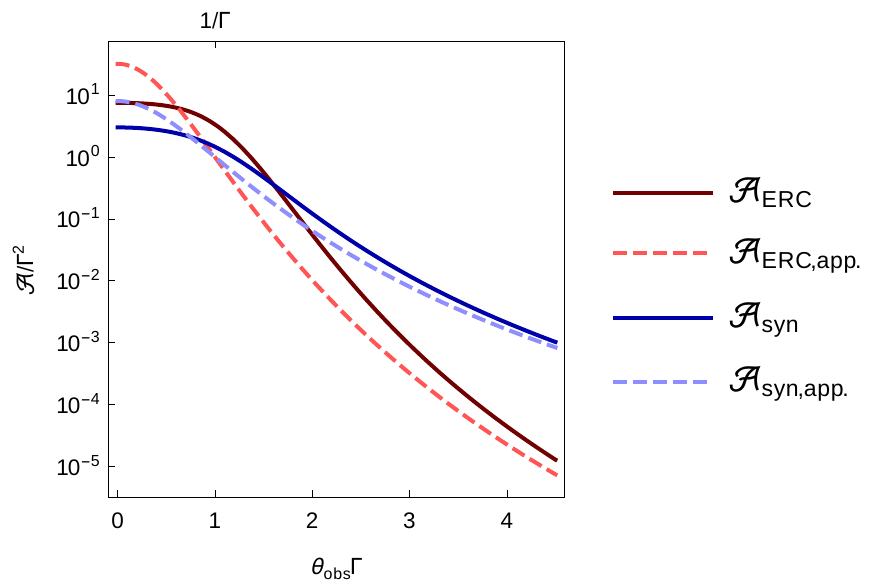}
    \caption{Luminosity amplification factors. Solid curves represent factors for a conical jet. Dashed curves represent aprroximate factors assuming that jet is linear.}
    \label{fig:amp-approximation} 
  \end{center}
\end{figure}

\bibliography{article}

\begin{thebibliography}{}
\makeatletter
\relax
\def\mn@urlcharsother{\let\do\@makeother \do\$\do\&\do\#\do\^\do\_\do\%\do\~}
\def\mn@doi{\begingroup\mn@urlcharsother \@ifnextchar [ {\mn@doi@}
  {\mn@doi@[]}}
\def\mn@doi@[#1]#2{\def\@tempa{#1}\ifx\@tempa\@empty \href
  {http://dx.doi.org/#2} {doi:#2}\else \href {http://dx.doi.org/#2} {#1}\fi
  \endgroup}
\def\mn@eprint#1#2{\mn@eprint@#1:#2::\@nil}
\def\mn@eprint@arXiv#1{\href {http://arxiv.org/abs/#1} {{\tt arXiv:#1}}}
\def\mn@eprint@dblp#1{\href {http://dblp.uni-trier.de/rec/bibtex/#1.xml}
  {dblp:#1}}
\def\mn@eprint@#1:#2:#3:#4\@nil{\def\@tempa {#1}\def\@tempb {#2}\def\@tempc
  {#3}\ifx \@tempc \@empty \let \@tempc \@tempb \let \@tempb \@tempa \fi \ifx
  \@tempb \@empty \def\@tempb {arXiv}\fi \@ifundefined
  {mn@eprint@\@tempb}{\@tempb:\@tempc}{\expandafter \expandafter \csname
  mn@eprint@\@tempb\endcsname \expandafter{\@tempc}}}

\bibitem[\protect\citeauthoryear{Ackermann et~al.,}{Ackermann
  et~al.}{2014}]{Ackermann2014}
Ackermann M.,  et~al., 2014, \apj, 786, 157

\bibitem[\protect\citeauthoryear{Asano \& Hayashida}{Asano \&
  Hayashida}{2015}]{Asano2015}
Asano K.,  Hayashida M.,  2015, \apjl, 808, L18

\bibitem[\protect\citeauthoryear{Begelman, Blandford  \& Rees}{Begelman
  et~al.}{1984}]{Begelman1984}
Begelman M.~C.,  Blandford R.~D.,   Rees M.~J.,  1984, \mn@doi [Rev. Mod.
  Phys.] {10.1103/RevModPhys.56.255}, 56, 255

\bibitem[\protect\citeauthoryear{Beskin \& Nokhrina}{Beskin \&
  Nokhrina}{2006}]{Beskin2006}
Beskin V.~S.,  Nokhrina E.~E.,  2006, \mn@doi [\mnras]
  {10.1111/j.1365-2966.2006.09957.x}, 367, 375

\bibitem[\protect\citeauthoryear{{Boccardi}, {Krichbaum}, {Bach}, {Mertens},
  {Ros}, {Alef}  \& {Zensus}}{{Boccardi} et~al.}{2015}]{Boccardi2015}
{Boccardi} B.,  {Krichbaum} T.~P.,  {Bach} U.,  {Mertens} F.,  {Ros} E.,
  {Alef} W.,   {Zensus} J.~A.,  2015, preprint, \href
  {http://adsabs.harvard.edu/abs/2015arXiv150906250B} {} (\mn@eprint {arXiv}
  {1509.06250})

\bibitem[\protect\citeauthoryear{Bogovalov \& Tsinganos}{Bogovalov \&
  Tsinganos}{2005}]{Bogovalov2005}
Bogovalov S.,  Tsinganos K.,  2005, \mn@doi [\mnras]
  {10.1111/j.1365-2966.2005.08671.x}, 357, 918

\bibitem[\protect\citeauthoryear{Bonning et~al.,}{Bonning
  et~al.}{2012}]{Bonning2012}
Bonning E.,  et~al., 2012, \apj, 756, 13

\bibitem[\protect\citeauthoryear{Bonnoli, Ghisellini, Foschini, Tavecchio  \&
  Ghirlanda}{Bonnoli et~al.}{2011}]{Bonnoli2011}
Bonnoli G.,  Ghisellini G.,  Foschini L.,  Tavecchio F.,   Ghirlanda G.,  2011,
  \mn@doi [\mnras] {10.1111/j.1365-2966.2010.17450.x}, 410, 368

\bibitem[\protect\citeauthoryear{{Capetti}, {Celotti}, {Chiaberge}, {de
  Ruiter}, {Fanti}, {Morganti}  \& {Parma}}{{Capetti}
  et~al.}{2002}]{2002A&A...383..104C}
{Capetti} A.,  {Celotti} A.,  {Chiaberge} M.,  {de Ruiter} H.~R.,  {Fanti} R.,
  {Morganti} R.,   {Parma} P.,  2002, \mn@doi [\aap]
  {10.1051/0004-6361:20011714}, \href
  {http://adsabs.harvard.edu/abs/2002A%26A...383..104C} {383, 104}

\bibitem[\protect\citeauthoryear{Carnerero et~al.,}{Carnerero
  et~al.}{2015}]{Carnerero2015}
Carnerero M.~I.,  et~al., 2015, \mn@doi [\mnras] {10.1093/mnras/stv823}, 450,
  2677

\bibitem[\protect\citeauthoryear{Celotti, Ghisellini  \& Chiaberge}{Celotti
  et~al.}{2001}]{Celotti2001}
Celotti A.,  Ghisellini G.,   Chiaberge M.,  2001, \mn@doi [\mnras]
  {10.1046/j.1365-8711.2001.04160.x}, 321, L1

\bibitem[\protect\citeauthoryear{{Chiaberge}, {Celotti}, {Capetti}  \&
  {Ghisellini}}{{Chiaberge} et~al.}{2000}]{2000A&A...358..104C}
{Chiaberge} M.,  {Celotti} A.,  {Capetti} A.,   {Ghisellini} G.,  2000, \aap,
  \href {http://adsabs.harvard.edu/abs/2000A%26A...358..104C} {358, 104}

\bibitem[\protect\citeauthoryear{Cohen, Romani, Filippenko, Cenko, Lott, Zheng
  \& Li}{Cohen et~al.}{2014}]{Cohen2014}
Cohen D.~P.,  Romani R.~W.,  Filippenko A.~V.,  Cenko S.~B.,  Lott B.,  Zheng
  W.,   Li W.,  2014, \apj, 797, 137

\bibitem[\protect\citeauthoryear{D'arcangelo et~al.,}{D'arcangelo
  et~al.}{2009}]{Darcangelo2009}
D'arcangelo F.~D.,  et~al., 2009, \apj, 697, 985

\bibitem[\protect\citeauthoryear{{Dermer}}{{Dermer}}{1995}]{Dermer1995}
{Dermer} C.~D.,  1995, \mn@doi [\apjl] {10.1086/187931}, 446, L63

\bibitem[\protect\citeauthoryear{Finke}{Finke}{2013}]{Finke2013}
Finke J.~D.,  2013, \apj, 763, 134

\bibitem[\protect\citeauthoryear{Ghisellini \& Madau}{Ghisellini \&
  Madau}{1996}]{Ghisellini1996}
Ghisellini G.,  Madau P.,  1996, \mn@doi [\mnras] {10.1093/mnras/280.1.67},
  280, 67

\bibitem[\protect\citeauthoryear{Ghisellini, Tavecchio  \&
  Chiaberge}{Ghisellini et~al.}{2005}]{Ghisellini2005}
Ghisellini G.,  Tavecchio F.,   Chiaberge M.,  2005, \mn@doi [A&A]
  {10.1051/0004-6361:20041404}, 432, 401

\bibitem[\protect\citeauthoryear{Ghisellini, Tavecchio, Maraschi, Celotti  \&
  Sbarrato}{Ghisellini et~al.}{2014}]{Ghisellini2014}
Ghisellini G.,  Tavecchio F.,  Maraschi L.,  Celotti A.,   Sbarrato T.,  2014,
  Nature, 515, 376

\bibitem[\protect\citeauthoryear{{Giommi, P.} et~al.,}{{Giommi, P.}
  et~al.}{2012}]{Giommi2012}
{Giommi, P.} et~al., 2012, \mn@doi [\aap] {10.1051/0004-6361/201117825}, 541,
  A160

\bibitem[\protect\citeauthoryear{{Giroletti} et~al.,}{{Giroletti}
  et~al.}{2004}]{2004ApJ...600..127G}
{Giroletti} M.,  et~al., 2004, \mn@doi [\apj] {10.1086/379663}, \href
  {http://adsabs.harvard.edu/abs/2004ApJ...600..127G} {600, 127}

\bibitem[\protect\citeauthoryear{Gracia, Tsinganos  \& Bogovalov}{Gracia
  et~al.}{2005}]{Gracia2005}
Gracia J.,  Tsinganos K.,   Bogovalov S.~V.,  2005, \mn@doi [\aap]
  {10.1051/0004-6361:200500175}, 442, L7

\bibitem[\protect\citeauthoryear{Hardee}{Hardee}{2007}]{Hardee2007a}
Hardee P.~E.,  2007, \apj, 664, 26

\bibitem[\protect\citeauthoryear{Hardee, Mizuno  \& Nishikawa}{Hardee
  et~al.}{2007}]{Hardee2007b}
Hardee P.,  Mizuno Y.,   Nishikawa K.-I.,  2007, \mn@doi [Astrophys. Space
  Sci.] {10.1007/s10509-007-9529-1}, 311, 281

\bibitem[\protect\citeauthoryear{Hayashida et~al.,}{Hayashida
  et~al.}{2015}]{Hayashida2015}
Hayashida M.,  et~al., 2015, \apj, 807, 79

\bibitem[\protect\citeauthoryear{Hovatta et~al.,}{Hovatta
  et~al.}{2014}]{Hovatta2014}
Hovatta T.,  et~al., 2014, \mn@doi [\mnras] {10.1093/mnras/stt2494}, 439, 690

\bibitem[\protect\citeauthoryear{{Janiak}, {Sikora}  \& {Moderski}}{{Janiak}
  et~al.}{2015a}]{Janiak2015}
{Janiak} M.,  {Sikora} M.,   {Moderski} R.,  2015a, preprint, \href
  {http://adsabs.harvard.edu/abs/2015arXiv150806500J} {} (\mn@eprint {arXiv}
  {1508.06500})

\bibitem[\protect\citeauthoryear{{Janiak}, {Sikora}  \& {Moderski}}{{Janiak}
  et~al.}{2015b}]{2015MNRAS.449..431J}
{Janiak} M.,  {Sikora} M.,   {Moderski} R.,  2015b, \mn@doi [\mnras]
  {10.1093/mnras/stv200}, \href
  {http://adsabs.harvard.edu/abs/2015MNRAS.449..431J} {449, 431}

\bibitem[\protect\citeauthoryear{{Jester}, {Harris}, {Marshall}  \&
  {Meisenheimer}}{{Jester} et~al.}{2006}]{2006ApJ...648..900J}
{Jester} S.,  {Harris} D.~E.,  {Marshall} H.~L.,   {Meisenheimer} K.,  2006,
  \mn@doi [\apj] {10.1086/505962}, \href
  {http://adsabs.harvard.edu/abs/2006ApJ...648..900J} {648, 900}

\bibitem[\protect\citeauthoryear{Kharb, Lister  \& Cooper}{Kharb
  et~al.}{2010}]{Kharb2010}
Kharb P.,  Lister M.~L.,   Cooper N.~J.,  2010, \apj, 710, 764

\bibitem[\protect\citeauthoryear{Komissarov, Vlahakis, Königl  \&
  Barkov}{Komissarov et~al.}{2009}]{Komissarov2009}
Komissarov S.~S.,  Vlahakis N.,  Königl A.,   Barkov M.~V.,  2009, \mn@doi
  [\mnras] {10.1111/j.1365-2966.2009.14410.x}, 394, 1182

\bibitem[\protect\citeauthoryear{Lister, Aller, Aller, Hovatta, Max-Moerbeck,
  Readhead, Richards  \& Ros}{Lister et~al.}{2015}]{Lister2015}
Lister M.~L.,  Aller M.~F.,  Aller H.~D.,  Hovatta T.,  Max-Moerbeck W.,
  Readhead A. C.~S.,  Richards J.~L.,   Ros E.,  2015, \apjl, 810, L9

\bibitem[\protect\citeauthoryear{Lyubarsky}{Lyubarsky}{2010}]{Lyubarsky2010}
Lyubarsky Y.~E.,  2010, \mn@doi [\mnras] {10.1111/j.1365-2966.2009.15877.x},
  402, 353

\bibitem[\protect\citeauthoryear{Marscher et~al.,}{Marscher
  et~al.}{2010}]{Marscher2010}
Marscher A.~P.,  et~al., 2010, \apjl, 710, L126

\bibitem[\protect\citeauthoryear{McKinney, Tchekhovskoy  \& Blandford}{McKinney
  et~al.}{2012}]{McKinney2012}
McKinney J.~C.,  Tchekhovskoy A.,   Blandford R.~D.,  2012, \mn@doi [\mnras]
  {10.1111/j.1365-2966.2012.21074.x}, 423, 3083

\bibitem[\protect\citeauthoryear{Meyer, Fossati, Georganopoulos  \&
  Lister}{Meyer et~al.}{2011}]{Meyer2011}
Meyer E.~T.,  Fossati G.,  Georganopoulos M.,   Lister M.~L.,  2011, \apj, 740,
  98

\bibitem[\protect\citeauthoryear{Mimica, Giannios, Metzger  \& Aloy}{Mimica
  et~al.}{2015}]{Mimica2015}
Mimica P.,  Giannios D.,  Metzger B.~D.,   Aloy M.~A.,  2015, \mn@doi [\mnras]
  {10.1093/mnras/stv825}, 450, 2824

\bibitem[\protect\citeauthoryear{Mizuno, Hardee  \& Nishikawa}{Mizuno
  et~al.}{2007}]{Mizuno2007}
Mizuno Y.,  Hardee P.,   Nishikawa K.-I.,  2007, \apj, 662, 835

\bibitem[\protect\citeauthoryear{{Mullin} \& {Hardcastle}}{{Mullin} \&
  {Hardcastle}}{2009}]{2009MNRAS.398.1989M}
{Mullin} L.~M.,  {Hardcastle} M.~J.,  2009, \mn@doi [\mnras]
  {10.1111/j.1365-2966.2009.15232.x}, \href
  {http://adsabs.harvard.edu/abs/2009MNRAS.398.1989M} {398, 1989}

\bibitem[\protect\citeauthoryear{{Nagai} et~al.,}{{Nagai}
  et~al.}{2014}]{2014ApJ...785...53N}
{Nagai} H.,  et~al., 2014, \mn@doi [\apj] {10.1088/0004-637X/785/1/53}, \href
  {http://adsabs.harvard.edu/abs/2014ApJ...785...53N} {785, 53}

\bibitem[\protect\citeauthoryear{Nalewajko \& Begelman}{Nalewajko \&
  Begelman}{2012}]{Nalewajko2012}
Nalewajko K.,  Begelman M.~C.,  2012, \mn@doi [\mnras]
  {10.1111/j.1365-2966.2012.22117.x}, 427, 2480

\bibitem[\protect\citeauthoryear{Nishikawa, Richardson, Koide, Shibata, Kudoh,
  Hardee  \& Fishman}{Nishikawa et~al.}{2005}]{Nishikawa2005}
Nishikawa K.-I.,  Richardson G.,  Koide S.,  Shibata K.,  Kudoh T.,  Hardee P.,
    Fishman G.~J.,  2005, \apj, 625, 60

\bibitem[\protect\citeauthoryear{{Piner} \& {Edwards}}{{Piner} \&
  {Edwards}}{2014}]{2014ApJ...797...25P}
{Piner} B.~G.,  {Edwards} P.~G.,  2014, \mn@doi [\apj]
  {10.1088/0004-637X/797/1/25}, \href
  {http://adsabs.harvard.edu/abs/2014ApJ...797...25P} {797, 25}

\bibitem[\protect\citeauthoryear{{Schwartz} et~al.,}{{Schwartz}
  et~al.}{2015}]{2015IAUS..313..219S}
{Schwartz} D.~A.,  et~al., 2015, in {Massaro} F.,  {Cheung} C.~C.,  {Lopez} E.,
    {Siemiginowska} A.,  eds,  IAU Symposium Vol. 313, IAU Symposium. pp
  219--224 (\mn@eprint {arXiv} {1505.06990}),
  \mn@doi{10.1017/S1743921315002215}

\bibitem[\protect\citeauthoryear{{Sikora}, {Stawarz}, {Moderski}, {Nalewajko}
  \& {Madejski}}{{Sikora} et~al.}{2009}]{2009ApJ...704...38S}
{Sikora} M.,  {Stawarz} {\L}.,  {Moderski} R.,  {Nalewajko} K.,   {Madejski}
  G.~M.,  2009, \mn@doi [\apj] {10.1088/0004-637X/704/1/38}, \href
  {http://adsabs.harvard.edu/abs/2009ApJ...704...38S} {704, 38}

\bibitem[\protect\citeauthoryear{{Sikora}, {Janiak}, {Nalewajko}, {Madejski}
  \& {Moderski}}{{Sikora} et~al.}{2013}]{Sikora2013}
{Sikora} M.,  {Janiak} M.,  {Nalewajko} K.,  {Madejski} G.~M.,   {Moderski} R.,
   2013, \mn@doi [\apj] {10.1088/0004-637X/779/1/68}, \href
  {http://adsabs.harvard.edu/abs/2013ApJ...779...68S} {779, 68}

\bibitem[\protect\citeauthoryear{Sol, Pelletier  \& Ass\'eo}{Sol
  et~al.}{1989}]{Sol1989}
Sol H.,  Pelletier G.,   Ass\'eo E.,  1989, \mn@doi [\mnras]
  {10.1093/mnras/237.2.411}, 237, 411

\bibitem[\protect\citeauthoryear{{Tavani}, {Vittorini}  \&
  {Cavaliere}}{{Tavani} et~al.}{2015}]{Tavani2015}
{Tavani} M.,  {Vittorini} V.,   {Cavaliere} A.,  2015, preprint, \href
  {http://adsabs.harvard.edu/abs/2015arXiv151006184T} {} (\mn@eprint {arXiv}
  {1510.06184})

\bibitem[\protect\citeauthoryear{Tavecchio \& Ghisellini}{Tavecchio \&
  Ghisellini}{2008}]{Tavecchio2008}
Tavecchio F.,  Ghisellini G.,  2008, \mn@doi [\mnras]
  {10.1111/j.1745-3933.2008.00441.x}, 385, L98

\bibitem[\protect\citeauthoryear{{Tavecchio} \& {Ghisellini}}{{Tavecchio} \&
  {Ghisellini}}{2015}]{2015arXiv150908710T}
{Tavecchio} F.,  {Ghisellini} G.,  2015, preprint, \href
  {http://adsabs.harvard.edu/abs/2015arXiv150908710T} {} (\mn@eprint {arXiv}
  {1509.08710})

\bibitem[\protect\citeauthoryear{Vercellone et~al.,}{Vercellone
  et~al.}{2011}]{Vercellone2011}
Vercellone S.,  et~al., 2011, \apjl, 736, L38

\bibitem[\protect\citeauthoryear{{Wardle} \& {Aaron}}{{Wardle} \&
  {Aaron}}{1997}]{1997MNRAS.286..425W}
{Wardle} J.~F.~C.,  {Aaron} S.~E.,  1997, \mn@doi [\mnras]
  {10.1093/mnras/286.2.425}, \href
  {http://adsabs.harvard.edu/abs/1997MNRAS.286..425W} {286, 425}

\bibitem[\protect\citeauthoryear{Wehrle et~al.,}{Wehrle
  et~al.}{1998}]{Wehrle1998}
Wehrle A.~E.,  et~al., 1998, \apj, 497, 178

\makeatother
\end{thebibliography}
\bibliographystyle{mnras}

\bsp	
\label{lastpage}
\end{document}